\documentclass[12pt,a4paper]{article}

\usepackage{amsmath,amssymb,amsfonts,amsthm,amscd}

\numberwithin{equation}{section}

\newcommand{\BbbR}{\mathbb{R}}
\newcommand{\BbbZ}{\mathbb{Z}}

\newcommand{\mdef}{\mathrel{\mathop:}=}  
 
\renewcommand{\d}{\mathrm{d}} 

\newcommand{\Z}{\mathbb{Z}}
\newcommand{\RPthree}{\mathbb{RP}^3}
\newcommand{\RPtwo}{\mathbb{RP}^2}
\newcommand{\sothree}{\mathrm{SO}(3)}

\newcommand{\sotwo}{\mathrm{SO}(2)}
\newcommand{\otwo}{\mathrm{O}(2)}
\newcommand{\sutwo}{\mathrm{SU}(2)}
\newcommand{\sun}{\mathrm{SU}(n)} 
\newcommand{\uone}{\mathrm{U}(1)}

\newcommand{\algsutwo}{\mathfrak{su}(2)}
\newcommand{\algsun}{\mathfrak{su}(n)} 
\newcommand{\alguone}{\mathfrak{u}(1)}

\newcommand{\M}{\mathcal{M}}
\newcommand{\ext}{_\mathrm{ext}}
\newcommand{\Aut}{\operatorname{Aut}}
\newcommand{\Id}{\operatorname{Id}}
\newcommand{\diag}{\operatorname{diag}}

\newcommand{\Genl}{G_{\mathrm{enl}}}
\newcommand{\Dhat}{{\hat D}}

\title{Topological geon black holes in\\[.5ex]
Einstein-Yang-Mills theory}
\author{
George T. Kottanattu\thanks{george.kottanattu@alumni.nottingham.ac.uk}
\ and
Jorma Louko\thanks{jorma.louko@nottingham.ac.uk}
\\
\noalign{\vspace{3ex}}
\small{\textit{School of Mathematical Sciences,
University of Nottingham,}}\\
\small{\textit{Nottingham NG7 2RD, UK}}
\\
\noalign{\vspace{3ex}}
\small{January 2010, revised September 2010}
\\
\noalign{\vspace{1ex}}
\small{Published in Commun.\ Math.\ Phys.\ 
\textbf{303} (2011) 127--148}
\\
\noalign{\vspace{1ex}}
\scriptsize{The final publication is available at www.springerlink.com}
}

\date{}

\begin{document}

\maketitle

\begin{abstract}
We construct topological geon quotients of 
two families of Einstein-Yang-Mills black holes. 
For K\"unzle's 
static, spherically symmetric
$\sun$ black holes 
with $n>2$, 
a geon quotient exists but generically requires
promoting 
charge conjugation into a gauge symmetry. 
For Kleihaus and Kunz's 
static, axially symmetric $\sutwo$ black holes 
a geon quotient exists 
without gauging charge conjugation, 
and the parity of the gauge field winding number determines 
whether the geon gauge bundle is trivial. 
The geon's gauge bundle structure 
is expected to have an imprint in the 
Hawking-Unruh effect 
for quantum
fields that couple to the 
background gauge field. 
\end{abstract}

\newpage

\section{Introduction}
\label{sec:intro}

Given a stationary black hole spacetime 
with a bifurcate Killing horizon, 
it may be possible to construct from it a 
time-orientable quotient spacetime in which 
the two exterior regions separated by the 
Killing horizon become identified. 
If the quotient is asymptotically flat, 
its spatial geometry is 
that of a compact manifold minus a point, 
with the omitted point at an asymptotically flat infinity. 
This makes the quotient a topological 
geon in the sense introduced by Sorkin~\cite{sorkin86}, 
as motivated by the earlier 
work in \cite{Sorkin:1979ja,friedmansorkin80,friedmansorkin82}. 
The showcase example is the $\RPthree$ geon 
\cite{Misner:1957mt,giulini-thesis,giulini-multiRP3s,Friedman:1993ty}, 
formed as a $\BbbZ_2$ quotient of Kruskal. 
There exist also quotients in which the infinity 
is only asymptotically locally flat,
and others in which the infinity is  
asymptotically anti-de~Sitter
or asymptotically locally anti-de~Sitter
\cite{Louko:1998hc,Louko:2000tp,Maldacena:2001kr,Louko:2004ej}. 
In this paper we shall understand topological 
geons to encompass all of these cases, 
the characteristic property being that the infinity 
consists of only one component. 

Topological geon black holes of the kind described above 
are unlikely to be created in an 
astrophysical star collapse, as their 
formation from conventional initial data would 
require a change in the spatial topology.
However, they provide an 
arena for 
the Hawking-Unruh effect in a setting where the 
black hole is eternal and has nonvanishing surface gravity 
but 
thermality for a quantum field cannot arise 
by the usual procedure of 
tracing over a  
causally disconnected exterior~\cite{birrell-davies}. 
There is still thermality, 
in the usual Hawking temperature, 
but only for a limited set of observations, 
and the non-thermal correlations bear an 
imprint of the absence of the 
causally disconnected exterior 
\cite{Louko:1998hc,Louko:2000tp,Maldacena:2001kr,%
Louko:2004ej,Louko:1998dj,Langlois:2004fv}. 
In a sense, the Hawking-Unruh effect 
on a topological geon black hole 
reveals to an exterior observer 
features of the geometry that are 
classically hidden behind the horizons. 
A recent review can be found in~\cite{Louko:2010tq}. 

When the black hole has a Maxwell field, 
it may be necessary to include charge 
conjugation in the map with 
which the black hole gauge bundle is 
quotiented into the geon gauge bundle~\cite{Louko:2004ej}. 
This happens for example for the
Reissner-Nordstr\"om hole, both with electric and magnetic charge; 
it also happens for the higher-dimensional Reissner-Nordstr\"om hole 
with electric charge in any dimension and with magnetic charge in even dimensions. 
Maxwell's theory on the geon must then incorporate 
charge conjugation as a gauge symmetry, 
rather than just as a global symmetry: 
technically, the gauge group is no longer 
$\uone \simeq \sotwo$ 
but
$\BbbZ_2 \ltimes \uone \simeq \otwo$, 
where the nontrivial element of 
$\BbbZ_2$ acts on $\uone$ by complex 
conjugation~\cite{Kiskis:1978ed}. 
The 
presence of the charge conjugation in the quotienting map 
can further be verified to leave its imprint in the 
Hawking-Unruh effect for a quantum field that couples to 
the background Maxwell field~\cite{bruschi-MGtalk,BruschiLouko}. 
By contrast, 
spherically symmetric Einstein-$\sutwo$ black holes  
admit a geon quotient without the inclusion of 
charge conjugation in the quotienting map, 
and the geon's gauge bundle 
is in fact trivial~\cite{Louko:2004ej}. 

The purpose of this paper is to construct two new 
families of
Einstein-Yang-Mills geon black holes. 
We shall specifically examine 
whether 
charge conjugation needs to be 
promoted into a gauge symmetry
when taking the geon quotient. 
We take the gauge group to be $\sun$ with $n\ge2$, 
a choice motivated physically by the appearance 
of these groups in particle physics and mathematically 
by their amenability to a unified treatment. 
We shall work with pure Einstein-Yang-Mills, 
but we note that these gauge groups, and the 
definition of spherical symmetry in terms 
of $\sutwo$ rather than $\sothree$
\cite{Kuenzle:1991wa,bartnik-struct}, 
provide opportunities for extensions that 
include spinor as well as scalar fields. 

In Sections 
\ref{sec:spherical-sun-hole}
and \ref{sec:geon-quotient}
we consider the static, spherically symmetric 
Einstein-$\sun$ black holes of K\"unzle
\cite{Kuenzle:1994ru} and their generalisations 
to a negative cosmological 
constant~\cite{Baxter:2007au,Baxter:2007at}. 
The case $n=2$ was covered in \cite{Louko:2004ej} 
as discussed above. 
For $n>2$ we show that a geon quotient 
exists and 
generically requires including 
charge conjugation in the quotienting map:  
the enlarged gauge group 
is 
$\BbbZ_2 \ltimes \sun$, 
where the nontrivial 
element of $\BbbZ_2$ acts on 
$\sun$ by complex conjugation. 
A quotient without charge conjugation 
is possible only for certain special field configurations, 
of which we give a complete list, 
and we show that the geon gauge bundle is then trivial. 

In Sections \ref{sec:stat-axially-symm}
and \ref{sec:su2quotient}
we consider the static, axially symmetric 
Einstein-$\sutwo$ black holes of 
Kleihaus and Kunz~\cite{Kleihaus:1997ic,Kleihaus:1997ws}. 
We show that all these holes admit a geon quotient without 
gauging charge conjugation. 
When the winding number of the 
gauge field configuration
is odd, 
the geon gauge bundle is trivial; 
this includes as a special case the 
spherically symmetric geon discussed in~\cite{Louko:2004ej}. 
When the winding number is even, 
the geon gauge bundle is nontrivial. 

Section \ref{sec:conclusions-outlook} 
summarises the results and 
discusses their relevance for the 
Hawking-Unruh effect.  

The metric signature is $({-}{+}{+}{+})$. 
Sections \ref{sec:spherical-sun-hole}
and \ref{sec:geon-quotient}
use the convention of an antihermitian gauge field, 
common in 
mathematical literature. 
Sections \ref{sec:stat-axially-symm}
and \ref{sec:su2quotient}
use the convention of a hermitian gauge field, 
common in physics literature. 
Homotopies are assumed smooth, 
without loss of generality~\cite{conlon-book}.

\section{Spherically symmetric $\sun$ black holes}
\label{sec:spherical-sun-hole}

In this section we review the relevant properties of the 
static, spherically symmetric
$\sun$ Einstein-Yang-Mills black holes of K\"unzle
\cite{Kuenzle:1994ru} and their generalisations 
to a negative cosmological 
constant~\cite{Baxter:2007au,Baxter:2007at}. 
We also give explicit Kruskal-like coordinates 
that extend these 
solutions across the Killing horizon. 
We assume $n>2$ when not 
explicitly mentioned otherwise, although 
most of the formulas hold also for 
$n=2$.

\subsection{Exterior ansatz}
\label{sec:descr-exter-spac}

Given a group action on the base space of a principal bundle, 
the notion of a symmetric gauge field can be formulated as invariance 
of the connection under an appropriate group action 
on the total space. 
Concretely, 
let $P$ be a principal bundle 
with base manifold~$\M$, 
projection $\pi: P \to \M$ 
and structure group $G$ with Lie algebra~$\mathfrak{g}$. 
Let $H$ be a group and $\phi: H \times \M \to \M$ 
its action on~$\M$. We say that the connection form 
$\omega \in \Lambda^1 (P, \mathfrak{g})$ 
is $H$-symmetric if the following three conditions hold: 
\begin{enumerate}
\item 
For each $h\in H$, there is a $\Phi_h \in \Aut(P)$ 
such that $\pi \circ \Phi_h = \phi_h \circ \pi$
with $\Phi_{\Id_H} = \Id_P$;  
\item 
$\Phi_h^* \omega  = \omega$ 
for all $h\in H$;   
\item 
The map $H \to \Aut(P)$ given by 
$h \mapsto \Phi_h$ is a group homomorphism. 
\end{enumerate}
This is essentially the 
definition adopted in~\cite{Harnad:1979in}. 
Condition 3 is known to have undesirably 
restrictive consequences in 
some situations, 
such as when $H$ is a translation group~\cite{molelekoa}, 
but for our applications 
the definition will be satisfactory. 

We take 
$\M$ to be a static, 
spherically symmetric spacetime 
and $G$ to be~$\sun$. 
In this subsection we specify $\M$ 
by a coordinate-based ansatz. 
The ansatz does not cover all regions of 
the Kruskal-type black and white whole 
spacetimes that will be introduced in 
Subsection~\ref{sec:krusk-like-extens}, 
but we shall see that the ensuing gauge field will 
remain spherically symmetric when appropriately 
continued beyond the coordinate 
singularities. 

The metric ansatz is 
\begin{equation}
\label{eq:1}
d s^2 = -N e^{-2\delta} \d t^2 + N^{-1} \d r^2 
+ r^2 (\d\theta^2 + \sin^2\!\theta\, \d\phi^2), 
\end{equation}
where the functions $N$ and $\delta$ depend only 
on the coordinate $r$ and we assume $N>0$ and $r>0$. 
The factor $\d\theta^2 + \sin^2\!\theta\, \d\phi^2$ is  
recognised as the metric on unit $S^2$, where 
$(\theta, \phi)$ are standard angle coordinates
with coordinate singularities at 
$\theta=0$ and $\theta=\pi$. 
It is evident from the ansatz 
that the metric has an $\sothree$ isometry 
whose orbits are spacelike with topology 
$S^2$ and area $4\pi r^2$. We refer to this 
$\sothree$ isometry as 
spherical symmetry. 

The metric \eqref{eq:1} is static, with 
the timelike Killing vector $\partial_t$ that is orthogonal 
to the hypersurfaces of constant~$t$. We refer to the coordinates 
$(t,r,\theta,\phi)$ as Schwarzschild-like coordinates 
and to $r$ as the area-radius. 
A~systematic derivation of the ansatz \eqref{eq:1} 
from the assumptions of spherical symmetry and staticity 
is given in~\cite{Kuenzle:1991wa}. 
The ansatz 
does not cover static, 
spherically symmetric spacetimes in which the area of the 
$\sothree$ orbits is constant (see 
Section 2
in~\cite{Kuenzle:1991wa}, 
Section 15.4 in~\cite{exactsols} or 
Exercise 32.1 in~\cite{MTW}), 
but this special case does not occur within the black hole spacetimes
in which we are interested.

The $\sothree$ action on $\M$ 
induces an action of the covering group  
$\sutwo$, by the double cover map 
$\sutwo \to \sutwo/\{\pm\Id\} \simeq \sothree$. 
Following 
\cite{Kuenzle:1991wa,bartnik-struct}, 
we adopt $\sutwo$ 
as the group $H$ of spherical symmetry in the 
above definition 
of a spherically symmetric connection. 
We shall now recall the resulting classification of 
these configurations 
and their description in an 
adapted Lie algebra basis~\cite{Kuenzle:1991wa}. 

The first part of the argument consists of determining all $\sun$
principal bundles that admit an $\sutwo$ action of the required
kind. For spacetimes that are regularly foliated 
by the $\sothree$ orbits, 
as is the case in~\eqref{eq:1}, 
this amounts to classifying all $\sun$ principal
bundles over~$S^2$. 
The classification relies on 
presenting $S^2$ as the quotient space $\simeq \sutwo/\uone$
of the base space and analysing the action of the isotropy subgroup 
$\uone \subset \sutwo$ on the total space of the bundle. 
The result is that, up to isomorphisms, the 
bundles are in one-to-one correspondence with the 
conjugacy classes of group homomorphisms from 
$\uone$ to $\sun$~\cite{Harnad:1979in}. 

A~convenient unique representative from each conjugacy class 
is the map $\lambda: \uone \to \sun,
\ z \mapsto \diag (z^{k_1},\ldots,z^{k_n})$, 
where the $n$ integers 
$k_1, \ldots, k_n$
satisfy $k_1 \ge k_2 \ge \cdots \ge k_n$ and sum to zero. 
It follows that 
the equivalence classes of
the $\sun$ principal bundles can be uniquely indexed by sets of $n$
integers $\{k_1, \ldots, k_n\}$ that sum to zero and are ordered so that 
$k_1 \ge k_2 \ge \cdots \ge k_n$. 

For the second part of the argument, one fixes the bundle and an
$\sutwo$ action on it and considers all connections that are invariant
under this action. 
Let the map 
$\lambda: \uone \to \sun$ be as defined above, 
and let $\lambda': \algsutwo \to \algsun$ denote the derivative
of $\lambda$ at the identity. 
A~theorem of
Wang \cite{wang58} 
then states that 
the invariant connections are in one-to-one correspondence with the set of 
linear maps $\Lambda: \algsutwo \to \algsun$ satisfying the conditions 
\begin{subequations}
\begin{align}
\label{eq:23}
\Lambda(X) &= \lambda'(X) , 
\\
\label{eq:22}
\Lambda \circ {\rm ad}_z &= {\rm ad}_{\lambda(z)} \circ \Lambda , 
\end{align}
\end{subequations}
for all $X \in \alguone$ and $z \in \uone$, 
where $\uone$ is again the
isotropy subgroup of the $\sutwo$ action. 
The curvature $F$ of these connections 
takes the form 
\begin{equation}
\label{eq:24}
  F(\tilde{X},\tilde{Y}) = \left[ \Lambda(X), \Lambda(Y) \right] -
  \Lambda\left(\left[X,Y\right]\right),
\end{equation}
where $X, Y \in \algsutwo$ and $\tilde{X}, \tilde{Y}$ are the
corresponding vector fields induced by the $\sutwo$ action on the
total space. 

We adopt for $\algsutwo$ the basis 
$\tau_l \mdef -\frac{i}{2}\sigma_l$, $l=1,2,3$, 
where $\sigma_l$ are the Pauli matrices, 
\begin{equation}
\label{eq:paulimatrices} 
  \sigma_1 = \begin{pmatrix} 0&1\\1&0 \end{pmatrix}, \ 
  \sigma_2 = \begin{pmatrix} 0&-i\\i&0 \end{pmatrix}, \ 
  \sigma_3 = \begin{pmatrix} 1&0\\0&-1 \end{pmatrix}. 
\end{equation}
We write $\Lambda_l \mdef \Lambda(\tau_l)$, 
$l=1,2,3$, and we may without loss of generality choose the 
isotropy subgroup $\uone$ to be embedded 
in $\sutwo$ as $z \mapsto
\left(\begin{smallmatrix}z &0\\ 0 &z^{-1}\end{smallmatrix}\right)$. 
From 
\eqref{eq:23} it then follows that $\Lambda_3 =
-\frac{i}{2} \diag\left( k_1,\ldots,k_n \right)$. The 
infinitesimal version of \eqref{eq:22} reads 
\begin{equation}
  \Lambda \left(\left[ \tau_3,\tau_l \right]\right) =
  \left[ \Lambda_3,\Lambda_l \right],
  \ \ i = 1,2, 
\end{equation}
which implies that  
$\Lambda_1$ and $\Lambda_2$ can be written as
\begin{equation}
\label{eq:26}
  \Lambda_1 = \frac{1}{2} \left( C-C^H \right) , 
  \quad \Lambda_2 =
  -\frac{i}{2} \left( C+C^H \right), 
\end{equation}
where $C$ is a strictly upper triangular complex $n\times n$ matrix, 
$C^H$ is its hermitian conjugate, and $C_{ij} \neq 0$ if and only if
$k_i = k_j + 2$. 

Evaluating \eqref{eq:24} on the $\algsutwo$ basis $\tau_l$ shows that 
the only non-vanishing component of the curvature form is 
$F(\tilde{\tau_1},\tilde{\tau_2}) = \left[ \Lambda_1,\Lambda_2
\right] - \Lambda_3$. As 
the base space $S^2$ is two-dimensional, 
the curvature form must be proportional to the 
spherically symmetric volume form 
$\sin\theta \, \d\theta \wedge \d\phi$. 
The curvature form on $S^2$ 
must hence take the form  
\begin{equation}
\label{eq:25}
  F =  \left(\left[ \Lambda_1,\Lambda_2 \right] - \Lambda_3 \right)
  \sin\theta \, \d\theta \wedge \d\phi . 
\end{equation}
A~corresponding connection form is 
\begin{equation}
\label{eq:Ahat-on-S2}
  \hat{A} \mdef \Lambda_1 \d\theta + \left(\Lambda_2 \sin\theta +
    \Lambda_3 \cos\theta \right) \d\phi. 
\end{equation}

Finally, the connection form $A$ on the 
four-dimensional spacetime \eqref{eq:1} can be decomposed as 
\begin{equation}
  A = \tilde{A} + \hat{A} , 
\end{equation}
where $\hat{A}$ is as in \eqref{eq:Ahat-on-S2} 
but the components of the matrix 
$C$ in
\eqref{eq:26} are allowed to depend on the coordinates $(t,r)$. 
The remaining part $\tilde{A}$ is an $\algsun$-valued 
one-form on the two-dimensional spacetime 
obtained by dropping the angles from~\eqref{eq:1}, 
invariant under the 
adjoint
action of the subgroup 
$\lambda ([ \uone ])$~\cite{Harnad:1979in}. 

In what follows we consider only the case 
\cite{Kuenzle:1994ru,Baxter:2007au,Baxter:2007at} where the
set of $n$ integers 
is $\left\{k_1,\ldots,k_n\right\}=\left\{n-1, n-3,
n-5, \ldots, -n+3, -n+1\right\}$. The connection form is taken to
have a vanishing Coulomb component, $A_t = 0$, and one can then choose the 
gauge so that also the radial component $A_r$ is zero. This means that 
we consider purely magnetic 
configurations of the form 
\begin{equation}
\label{eq:3}
A = \Lambda_1 \d\theta + \left( \Lambda_2 \sin\theta + \Lambda_3
\cos\theta \right) \d\phi , 
\end{equation}
where the traceless antihermitian matrices 
$\Lambda_1$, 
$\Lambda_2$ and $\Lambda_3$ are given by 
\begin{subequations}
\label{eq:Lambdas}
\begin{align}
\label{eq:Lambda1}
  \Lambda_1 &= \frac{1}{2}
  \begin{pmatrix}
    0& w_1& \\
    -w_1& 0& w_2& \\
    & -w_2& 0& w_3& \\
    &&\ldots&\ldots&\ldots\\
    &&&-w_{n-2}& 0& w_{n-1}\\
    &&&& -w_{n-1}& 0
  \end{pmatrix} , 
\end{align}
\begin{align}
\label{eq:Lambda2}
  \Lambda_2 &= -\frac{i}{2}
  \begin{pmatrix}
    0& w_1& \\
    w_1& 0& w_2& \\
    &w_2& 0& w_3& \\
    &&\ldots&\ldots&\ldots\\
    &&&w_{n-2}& 0& w_{n-1}\\
    &&&&w_{n-1}& 0
  \end{pmatrix} , 
\end{align}
\begin{align}
\label{eq:Lambda3}
  \Lambda_3 &= -\frac{i}{2}
  \begin{pmatrix}
    n-1\\
    &n-3\\
    &&n-5\\
    &&&\ldots\\
    &&&&-n+3\\
    &&&&&-n+1
  \end{pmatrix}, 
\end{align}
\end{subequations}
and the real-valued functions~$w_j$, $j=1,\ldots,n-1$, 
depend only on the
coordinate~$r$. 

We end the subsection with three comments. 

First, the one-form 
\eqref{eq:3} has a Dirac string singularity as $\theta \to
0$ and $\theta \to \pi$~\cite{Volkov:1998cc}. 
The regularity of the 
curvature form \eqref{eq:25} shows that 
this singularity is a gauge artefact. 
As the triviality of the fundamental group of
$\sun$ implies that $\sun$ principal bundles over two-spheres are
trivial \cite{steenrod51,naber97:topol}, the one-form \eqref{eq:3} must 
therefore be a local representative of a 
connection one-form in the trivial $\sun$ bundle over the spacetime. 
We shall explicitly remove the Dirac string singularity 
in Subsection~\ref{sec:remov-dirac-string}. 

Second, 
the ansatz \eqref{eq:3} has a residual gauge freedom 
in that a gauge transformation by 
\begin{align}
\label{eq:wsigntransf}
e^ {ik\pi/n}
\diag\bigl( \, 
\underbrace{-1,-1,\ldots,-1}_{k},
\underbrace{1,1,\ldots,1}_{n-k} \, \bigr)
\in \sun
\end{align}
leaves $w_j$ invariant for $j\ne k$ 
but changes the sign of $w_k$ 
\cite{Kuenzle:1994ru}. 
We shall use this gauge 
freedom to simplify the special 
geon configurations that 
will be found in Subsection~\ref{sec:spher-specsols}. 

Third, we note that the embedding 
of the isotropy subgroup 
$\uone\subset \sutwo$ in $\sun$ maps 
$\exp(2\pi\tau_3) = -\Id_{\sutwo}$  
to $\exp\bigl(2\pi \Lambda_3\bigr)$, 
which for our configurations 
\eqref{eq:Lambdas} equals 
$\Id_{\sun}$ for odd $n$ and $-\Id_{\sun}$ for even~$n$. 
A~gauge transformation by 
$\exp\bigl(2\pi \Lambda_3\bigr)$ 
hence leaves the ansatz \eqref{eq:3} invariant, 
and by the discussion in 
Subsection \ref{sec:remov-dirac-string} the same 
holds also in a globally regular gauge 
in which the Dirac string singularities of 
\eqref{eq:3} have been removed. 
For the gauge configurations that we consider, 
the spherical symmetry action of 
$\sutwo$ hence projects to a 
spherical symmetry action of
$\sutwo/\{\pm\Id\} \simeq \sothree$. 

\subsection{Nondegenerate Killing horizon: Kruskal-like extension}
\label{sec:krusk-like-extens}

The metric \eqref{eq:1} and the connection form \eqref{eq:3}
give an ansatz that can be inserted in the Einstein-Yang-Mills field 
equations. We are interested in spacetimes that have 
a nondegenerate Killing horizon at $r=r_h>0$, where 
$N(r_h)=0$ and $N'(r_h)>0$, the prime indicating 
derivative with respect to~$r$.  
Initial data for integrating the field equations from $r=r_h$ towards 
increasing $r$ then consists of~$r_h$, $\delta(r_h)$ and 
$w_j(r_h)$, $j=1, \ldots, n-1$. 
Local solutions in some neighbourhood of the horizon exist under a weak 
regularity restriction 
on~$w_j(r_h)$ \cite{Kuenzle:1994ru,Baxter:2007au}. 
Not all of these local solutions extend to an asymptotically 
flat (for a vanishing cosmological constant) 
or asymptotically anti--de~Sitter (for a negative cosmological constant) 
infinity at $r\to\infty$, 
but for those that do, the solution is a static region of a nondegenerate 
black hole spacetime. 
Numerical results are given in 
\cite{Kuenzle:1994ru,Baxter:2007au,Baxter:2007at,Kleihaus:1997rb}.  

To extend the metric across the Killing horizon, 
we start in the exterior region 
and define the 
Kruskal-like coordinates 
$(U,V,\theta,\phi)$ by
\begin{subequations}
\label{Kruskal-type}
\begin{align}
  U &\mdef -\exp \left[ -\alpha \left(t - \int_{r_0}^r
      \frac{e^{\delta(r)}}{N(r)} \d r \right)\right] , \\
  V &\mdef \exp \left[ \alpha \left( t + \int_{r_0}^r
      \frac{e^{\delta(r)}}{N(r)} \d r \right)\right] , 
\end{align}
\end{subequations}
where $\alpha := \frac12 N'(r_h) e^{-\delta(r_h)}$ and the constant 
$r_0$ is chosen so that 
the product~$UV$, 
\begin{equation}
\label{eq:UV-def}
  UV = 
  -\exp\left[2\alpha\int_{r_0}^r \frac{e^{\delta(r)}}{N(r)}\d r\right],
\end{equation}
has the Taylor expansion 
\begin{equation}
\label{expansionUV}
  UV = -\frac{r-r_h}{r_h}
  \left[1+\left(\delta'(r_h)
      -\frac{1}{2}\frac{N''(r_h)}{N'(r_h)}\right)\left(r-r_h\right) 
    +O\left(\left(r-r_h\right)^2\right)\right] 
\end{equation}
as $r \to r_h$. 
It follows that in the exterior we have $U < 0$ and $V>0$, 
and the Killing horizon is at 
$UV\to0_-$. Whether $UV$ is bounded below depends on the 
asymptotic behaviour of the metric at large~$r$, 
but this will not affect what follows. 

The metric in the coordinates $(U,V,\theta,\phi)$ reads 
\begin{equation}
\label{metricUV}
d s^2 = \frac{1}{\alpha^2}\frac{N(r) e^{-2\delta(r)}}{UV}\d U \d V +
r^2 (\d\theta^2+\sin^2\!\theta\,\d\phi^2),
\end{equation}
where $r$ is a function of 
$UV$ via~\eqref{eq:UV-def}. 
Inverting \eqref{expansionUV} as 
\begin{equation}
\frac{r-r_h}{r_h}  = - UV \left[ 1 +
\left(\delta'(r_h)-\frac{1}{2}\frac{N''(r_h)}{N'(r_h)}\right) r_h UV
+O\left(\left(UV\right)^2\right)\right], 
\end{equation}
we find that the metric \eqref{metricUV} has the near-horizon expansion 
\begin{align}
d s^2 &= -\frac{4r_h}{N'(r_h)}\left[1
+\left(3\delta'(r_h)-\frac{N''(r_h)}{N'(r_h)}\right) r_h UV +
O\left(\left(UV\right)^2\right)\right]\d U\d V 
\notag
\\[1ex]
& \hspace{5ex}
+ r^2 \left( \d\theta^2+\sin^2\!\theta\,\d\phi^2 \right), 
\label{metricUVexpanded}
\end{align}
which is regular across $UV=0$. 
The metric can hence be extended 
from the original, `right-hand-side' 
exterior to the black hole interior where $U>0$ and $V>0$, to the 
white hole interior where $U<0$ and $V<0$ and 
to the `left-hand-side' exterior where $U > 0$ and $V<0$. If the functions 
$N(r)$ and $\delta(r)$ are smooth at $r=r_h$, 
it further follows that the metric in the Kruskal coordinates 
is smooth at the horizon. 
Whether $UV$ has an upper limit 
in the black and white hole regions, and whether 
there are further Killing horizons past these regions, 
will not affect what follows. 

The extension of the gauge potential across the horizon 
is given by \eqref{eq:3} and \eqref{eq:Lambdas}, with 
$w_j = w_j\bigl(r(UV)\bigr)$. The extension is regular 
since $w_j(r_h)$ are part of the boundary 
data for the exterior solution, 
and the extension is smooth if 
$w_j(r)$ are smooth at $r=r_h$. 

The resulting Kruskal-like extension is 
spherically symmetric, 
with the orbits of the $\sothree$ 
isometry being spacelike and having topology~$S^2$. 
As the exterior gauge potential ansatz 
\eqref{eq:3}
does not have terms proportional to $\d t$ or~$\d r$, and 
as the coefficients depend only on~$r$, 
the gauge field on the Kruskal-like extension 
is spherically symmetric in  
the same sense as in the exterior.

\section{Geon quotient of the spherically 
symmetric $\sun$ black hole}
\label{sec:geon-quotient}

We wish to take a geon quotient of the Kruskal-like $\sun$ black hole
of Section~\ref{sec:spherical-sun-hole}.  For the spacetime manifold
this is a straightforward adaptation of the procedure with which the
Kruskal manifold is quotiented into the $\RPthree$ geon
\cite{Misner:1957mt,giulini-thesis,giulini-multiRP3s,Friedman:1993ty},
and we shall review the requisite notions in
Subsection~\ref{sec:spacetime-quotient}. The new issues arise with
including in the quotient the principal bundle in which the gauge
field lives. These issues will be addressed in Subsections 
\ref{sec:spher-specsols}--\ref{sec:spher-gensol-enlarged}.

For presentational simplicity, 
we take the gauge group of the 
black hole bundle 
to be $\sun$ for odd $n$ and $\sun/\{\pm\Id\}$ for even~$n$. 
We denote this gauge group by~$G$. 
We write equations in $G$ as matrix equations 
in the defining matrix representation, 
understanding for even 
$n$ the matrices to be defined up to overall sign. 
Proceeding with the gauge group $\sun$ for all 
$n$ would yield the same end results but we shall see in 
Subsection \ref{sec:remov-dirac-string} that our choice of 
$G$ will shorten the technical steps.

\subsection{Geon quotient of the spacetime manifold}
\label{sec:spacetime-quotient}

Let $\M$ denote the spacetime manifold of the Kruskal-like extension,
with the metric constructed in
Subsection~\ref{sec:krusk-like-extens}. $\M$~is covered by the
Kruskal-like coordinates $(U,V,\theta,\phi)$, with the usual
coordinate singularities of the angle coordinates at $\theta=0$ and
$\theta=\pi$. In addition to the three Killing vectors of the
$\sothree$ isometry, $\M$ has the Killing vector $\xi := V\partial_V -
U\partial_U$, which is timelike in the right and left exterior
quadrants where $UV<0$, spacelike in the black and white hole
quadrants where $UV>0$ and null on the bifurcate Killing horizon
$UV=0$. In the right exterior quadrant where $V>0$ and $U<0$, covered
by the metric~\eqref{eq:1}, $\xi = \alpha^{-1} \partial_t$.

$\M$ has topology $\BbbR^2 \times S^2$ and is foliated by
spacelike hypersurfaces of topology $\BbbR \times S^2 \simeq S^3
\setminus \{\text{two points}\}$, each omitted point being at a
spatial infinity. The product $UV$ may be bounded below by some
negative constant, depending on the nature of the spatial
infinities~\cite{Louko:2004ej}, and it may be bounded above by some
positive constant, depending on the properties of the spacetime in the
black and white hole regions. The possible existence of such bounds will
not affect what follows.

Consider now the map 
\begin{align}
J: \M &\to \M ; 
\notag\\
(U,V,\theta,\phi) & \mapsto (V,U,\pi-\theta,\phi+\pi) , 
\label{eq:Jspher-def}
\end{align}
where the action on the angle coordinates is recognised as the $S^2$
antipodal map and is understood in this sense at the coordinate
singularities. $J$~is an involutive isometry without fixed points,
and it preserves both space and time orientation. The quotient $\M' :=
\M / \{\Id, J\}$ is therefore a time and space orientable
spacetime. $\M'$~is foliated by spacelike hypersurfaces of topology
$\left(S^3 \setminus \{\text{two points}\}\right) \! /\BbbZ_2 \simeq
\RPthree \setminus \{\text{point}\}$, with the omitted point being at
a spatial infinity. 
As recalled in Section~\ref{sec:intro}, 
these properties make $\M'$ a topological geon spacetime,
in the asymptotically flat case in the sense of Sorkin
\cite{sorkin86,Sorkin:1979ja,friedmansorkin80,friedmansorkin82} and in
the asymptotically anti-de~Sitter case in the generalised sense of
\cite{Louko:1998hc,Louko:2000tp,Maldacena:2001kr,Louko:2004ej}.

As the quotienting identifies the two exterior regions of~$\M$,
$\M'$ is an eternal black and white hole spacetime, with a single
exterior region that isometric to one exterior region of~$\M$. We may
hence refer to $\M'$ as a topological geon black hole. The conformal
diagram depends on the character of the spatial infinity and on
the structure of the black hole interior: representative samples may
be found in 
\cite{Friedman:1993ty,Louko:1998hc,Louko:1998dj,Louko:2010tq}. 

We end with two observations on the isometries of~$\M'$. 

First, as the $\sothree$ action on $\M$ commutes with~$J$, there is an
induced $\sothree$ action on~$\M'$, with two-dimensional spacelike
orbits. The generic orbits have again topology~$S^2$, but the special
orbits that come from the $U=V$ subset of $\M$ have
topology~$\RPtwo$. We shall regard $\M'$ as a spherically
symmetric spacetime despite these exceptional orbits. 

Second, $J$ changes the sign of the Killing vector~$\xi$. The
isometries generated by $\xi$ on $\M$ do therefore not induce an
isometry on $\M'$: while such isometries exist within the 
exterior region of~$\M'$, they cannot be extended past 
the horizon. The ramifications of this phenomenon for the
Hawking-Unruh effect on related Einstein(-Maxwell) topological geon
black holes have been investigated in
\cite{Louko:1998hc,Louko:2000tp,Maldacena:2001kr,Louko:1998dj,%
Langlois:2004fv,bruschi-MGtalk,BruschiLouko}.

\subsection{Special gauge field configurations: 
geon quotient of the principal bundle}
\label{sec:spher-specsols}

We now embark on the task of examining whether the spacetime quotient
$\M \mapsto \M'$ can be extended to the principal bundle in which the
gauge field lives.

Let $A\ext$ denote the gauge potential 
\eqref{eq:3} on~$\M$, 
\begin{align}
\label{eq:7}
A\ext \mdef\Lambda_1 \d\theta + \left( \Lambda_2 \sin\theta + \Lambda_3
\cos\theta \right) \d\phi . 
\end{align}
We need to examine whether there is a bundle map that projects to $J$
on $\M$ and leaves the gauge field invariant. In terms of the
gauge potential $A\ext$~\eqref{eq:7}, this amounts to asking whether $J$ maps $A\ext$
to a gauge-equivalent gauge potential.  Denoting by $A\ext^J$ the
pull-back of $A\ext$ by~$J$, we thus seek a gauge function $\Omega: \M
\to G$, such that a gauge transformation by $\Omega$ maps $A\ext^J$
back to $A\ext$,
\begin{equation}
\label{eq:9}
\Omega A\ext^J \Omega^{-1} + \Omega \d \Omega^{-1} = A\ext.
\end{equation}
From \eqref{eq:Jspher-def} we find 
\begin{equation}
\label{eq:8}
A\ext^J 
= -\Lambda_1 \d\theta + \left( \Lambda_2
    \sin\theta - \Lambda_3 \cos\theta \right) \d\phi . 
\end{equation}
As neither \eqref{eq:7} nor \eqref{eq:8} involves 
$\d U$ or~$\d V$, we may assume $\Omega$ 
to depend only on the angular coordinates $(\theta,\phi)$. 
Equation \eqref{eq:9} is then equivalent to the pair 
\begin{subequations}
\label{eq:9broken} 
\begin{align}
  - \Omega \Lambda_1 \Omega^{-1} + \Omega \partial_\theta
  \Omega^{-1} &= \Lambda_1 , 
  \label{eq:12} 
  \\
  \Omega(\Lambda_2 \sin\theta - \Lambda_3 \cos\theta)\Omega^{-1}
  + \Omega \partial_\phi \Omega^{-1} &= \Lambda_2 \sin\theta +
  \Lambda_3 \cos\theta. 
  \label{eq:13}
\end{align}
\end{subequations}

To find a necessary condition for a solution to \eqref{eq:9broken} to
exist, we consider the field strengths of $A\ext$ and~$A\ext^J$. These
can be computed from
\begin{equation}
F(X,Y) = \d A(X,Y) + \frac{1}{2}[A(X),A(Y)],
\end{equation}
with the result 
\begin{subequations}
\label{eq:FandFJ}
\begin{align}
  F\ext &= \partial_U \Lambda_1 \d U\wedge\d\theta +
  \partial_V\Lambda_1 \d V\wedge\d\theta 
  + \partial_U\Lambda_2\sin\theta \, \d U\wedge\d\phi
  \notag
  \\
  &\hspace{3ex} 
  +\partial_V\Lambda_2\sin\theta \d V\wedge\d\phi +
  \left([\Lambda_1,\Lambda_2]-\Lambda_3\right)\sin\theta
  \, \d\theta\wedge\d\phi,
  \label{eq:10}
  \\[1ex]
  F\ext^J &= -\partial_U \Lambda_1 \d U\wedge\d\theta -
  \partial_V\Lambda_1 \d V\wedge\d\theta 
  + \partial_U\Lambda_2\sin\theta \, \d U\wedge\d\phi
  \notag
  \\
  &\hspace{3ex} + \partial_V\Lambda_2\sin\theta \d V\wedge\d\phi -
  \left([\Lambda_1,\Lambda_2]-\Lambda_3\right)\sin\theta
  \, \d\theta\wedge\d\phi.
  \label{eq:11}
\end{align}
\end{subequations}
From \eqref{eq:9} it follows that these 
field strengths are related by 
\begin{equation}
  \Omega F\ext^J \Omega^{-1} =  F\ext.
  \label{eq:OFO=F}
\end{equation}
Inserting \eqref{eq:FandFJ} in \eqref{eq:OFO=F} 
and using the fact that $\Omega$ only 
depends on the angular coordinates, 
\eqref{eq:OFO=F} 
reduces to 
\begin{subequations}
\label{eq:14all}
\begin{align}
  \Omega \Lambda_1 \Omega^{-1} &= -\Lambda_1 , 
  \label{eq:14p}\\
  \Omega \Lambda_2 \Omega^{-1} &= \Lambda_2 , 
  \label{eq:14pp}\\
  \Omega ([\Lambda_1,\Lambda_2]-\Lambda_3) \Omega^{-1} &= -
  ([\Lambda_1,\Lambda_2]-\Lambda_3). \label{eq:14}
\end{align}
\end{subequations}
Simplifying \eqref{eq:14} with the help of 
\eqref{eq:14p} and \eqref{eq:14pp} shows that 
the set \eqref{eq:14all}
is equivalent to
\begin{subequations}
\label{conditions_genericsols}
\begin{align}
  \Omega \Lambda_1 \Omega^{-1} &= -\Lambda_1, \label{eq:17p}\\
  \Omega \Lambda_2 \Omega^{-1} &= \Lambda_2, \label{eq:17}\\
  \Omega \Lambda_3 \Omega^{-1} &= -\Lambda_3. \label{eq:15}
\end{align}
\end{subequations}
The set \eqref{conditions_genericsols} is hence a necessary
condition for \eqref{eq:9broken} to hold. 

To analyse~\eqref{conditions_genericsols}, 
observe first from \eqref{eq:Lambda3} that 
$\Lambda_3$ and $-\Lambda_3$ are diagonal and their diagonal 
elements appear in the reverse order, 
\begin{equation}
  -\Lambda_3 = -\frac{i}{2}
  \begin{pmatrix}
    -n+1\\
    &-n+3\\
    &&-n+5\\
    &&&\cdots\\
    &&&&n-3\\
    &&&&&n-1
  \end{pmatrix}.
\end{equation}
Identity \eqref{eq:15} thus implies that $\Omega$ has the form
\begin{equation} 
\label{eq:Omega-alphas}
\Omega =  (-i)^{n-1}
  \begin{pmatrix}
    &&&&\alpha_1\\
    &&&\alpha_2\\
    &&\cdots\\
    &\alpha_{n-1}\\
    \alpha_n
  \end{pmatrix}
\end{equation}
where $\alpha_j$ are complex numbers with unit 
magnitude and $\prod_{j=1}^n \alpha_j =1$. 


Consider then 
\eqref{eq:17p} and~\eqref{eq:17}. 
Using $\Omega^{-1} = \overline{\Omega}^T$, 
where the overline denotes complex conjugation and 
$^T$ transposition, we find 
\begin{align}
  \Omega \Lambda_1 \Omega^{-1} = \frac{1}{2}
  \begin{pmatrix}
    0 &-\alpha_1\overline{\alpha}_2 w_{n-1}\\
    \alpha_2\overline{\alpha}_1 w_{n-1} &0 &-\alpha_2\overline{\alpha}_3 w_{n-2} \\
    &\cdots &\cdots &\cdots \\
    & &\alpha_{n-1}\overline{\alpha}_{n-2} w_2 &0 &-\alpha_{n-1}\overline{\alpha}_n w_1\\
    & & &\alpha_n\overline{\alpha}_{n-1} w_1 &0
  \end{pmatrix}.
\label{eq:OL1Oinv}
\end{align}
By \eqref{eq:Lambda1} and~\eqref{eq:OL1Oinv}, 
\eqref{eq:17p} reduces to 
the set 
\begin{align}
\alpha_1 \overline{\alpha}_2 w_{n-1} 
&= w_1 = \alpha_2 \overline{\alpha}_1 w_{n-1}, 
\notag
\\
\alpha_2 \overline{\alpha}_3 w_{n-2} 
&= w_2 = \alpha_3 \overline{\alpha}_2 w_{n-2}, 
\notag
\\
& \hspace{3.5ex} \vdots
\notag
\\
\alpha_{n-2} \overline{\alpha}_{n-1} w_2 
&= w_{n-2} = \alpha_{n-1} \overline{\alpha}_{n-2} w_2, 
\notag
\\
\alpha_{n-1} \overline{\alpha}_n w_1 
&= w_{n-1} = \alpha_n \overline{\alpha}_{n-1} w_1, 
\label{eq:coupled-alphaw}
\end{align}
and it can be similarly verified that
also \eqref{eq:17} reduces to~\eqref{eq:coupled-alphaw}. 

As $n>2$ by assumption, 
alphas satisfying \eqref{eq:coupled-alphaw}
do not exist for generic gauge field configurations. 
There is however a special class of gauge field 
configurations for which such alphas exist. 
If $w_j$ is vanishing, the $j$th line of 
\eqref{eq:coupled-alphaw} requires $w_{n-j}$ to vanish. 
If $w_j$ is nonvanishing, the $j$th line of 
\eqref{eq:coupled-alphaw} implies 
$w_{n-j} = \epsilon_j w_j$ and 
$\alpha_{j+1} = \epsilon_j \alpha_j$, 
where $\epsilon_j \in \{-1,+1\}$, 
and if $n$ is even, 
$\epsilon_{n/2} = 1$. 
A~necessary condition for the alphas to exist 
is therefore that the gauge potential functions satisfy 
$w_{n-j} = \epsilon_j w_j$ for all~$j$, with 
$\epsilon_j \in \{-1,+1\}$ 
and $\epsilon_j = \epsilon_{n-j}$. 
Note that this condition is 
compatible with the 
radial evolution equation  
for the gauge potential functions~\cite{Kuenzle:1994ru}. 
As observed in Subsection~\ref{sec:descr-exter-spac}, 
the sign of each $w_j$ can be independently 
changed by a gauge transformation. 
The gauge can therefore be chosen 
so that the 
necessary condition for the 
alphas to exist reads 
\begin{align}
w_{n-j} = w_j, 
\hspace{3ex}
\forall j . 
\label{eq:nessGcondition} 
\end{align}

Now, when \eqref{eq:nessGcondition} holds, 
\eqref{eq:coupled-alphaw} is solved by $\alpha_j=1 \ \forall j$, 
and $\Omega$ \eqref{eq:Omega-alphas} then takes the form 
\begin{equation} 
\label{eq:Omega-alphas-solved}
\Omega =  (-i)^{n-1}
  \begin{pmatrix}
    &&&&1\\
    &&&1\\
    &&\cdots\\
    &1\\
    1
  \end{pmatrix} . 
\end{equation}
The condition \eqref{eq:nessGcondition} is hence also sufficient for a
solution to \eqref{conditions_genericsols} to exist, and 
as \eqref{eq:Omega-alphas-solved} is independent of the angles, 
this also provides a solution to~\eqref{eq:9broken}. 

We summarise. 
The necessary and sufficient condition for a geon quotient with 
the gauge group $G$ to exist is~\eqref{eq:nessGcondition}, 
up to gauge transformations. 
When \eqref{eq:nessGcondition} holds, 
the gauge transformation 
that compensates for $J$ in the quotienting bundle 
map is given by~\eqref{eq:Omega-alphas-solved}. 

We note in passing that for 
$n=2$ the only gauge potential function is $w_1$ 
and the equations \eqref{eq:coupled-alphaw} 
have the solution $\alpha_1 = \alpha_2 = 1$. 
This yields the purely magnetic special case of the 
$\sutwo$ geon described in~\cite{Louko:2004ej}.

\subsection{Triviality of the black hole bundle}
\label{sec:remov-dirac-string}

Up to now we have been working in a gauge in which the gauge 
potential $A\ext$ \eqref{eq:7} has Dirac string singularities at
$\theta=0$ and $\theta=\pi$. 
As noted at the end of Subsection~\ref{sec:descr-exter-spac}, 
the gauge bundle over the 
Kruskal-like spacetime $\M$ is trivial, 
and a globally regular gauge on $\M$ must hence exist. 
In this subsection we transform $A\ext$ into a globally regular gauge. 
This will be used in Subsection \ref{sec:disc-non-triv}
to analyse the gauge bundle over the geon spacetime. 

To begin, observe that gauge transformations by the functions 
\begin{subequations}
\label{eq:OmegaNS}
\begin{align}
\label{eq:OmegaN}
  \Omega_N &\mdef \diag \bigl( e^{-i(n-1)\phi/2}, e^{-i(n-3)\phi/2} ,
  \ldots, e^{-i(-n+3)\phi/2} , e^{-i(-n+1)\phi/2} \bigr) , 
  \\
\label{eq:OmegaS}
  \Omega_S &\mdef \diag \bigl( e^{i(n-1)\phi/2}, e^{i(n-3)\phi/2} ,
  \ldots, e^{i(-n+3)\phi/2} , e^{i(-n+1)\phi/2} \bigr) , 
\end{align} 
\end{subequations}
make the gauge potential $A\ext$ \eqref{eq:7} regular everywhere except 
respectively at $\theta=\pi$ and $\theta=0$. 
This is the step where taking the gauge group to be 
$\sun/\{\pm\Id\}$ for even $n$ shortens the discussion, 
as the expressions \eqref{eq:OmegaNS} are not 
single-valued in $\sun$ for even~$n$. 

It therefore suffices to find a gauge function  
$H: S^2 \setminus 
\left( 
\{ \theta=0 \} \cup \{ \theta=\pi \}
\right)$ 
that agrees with $\Omega_N$ in some punctured 
neighbourhood of $\theta=0$, 
agrees with $\Omega_S$ in some punctured 
neighbourhood of $\theta=\pi$, and interpolates in between: 
a transformation by $H$ puts 
$A\ext$ \eqref{eq:7} into a globally regular gauge. 
We shall show that such gauge functions exist. 

Let first $n$ be odd. The formulas \eqref{eq:OmegaNS} for 
$\Omega_N$ and $\Omega_S$ 
define two paths in $G=\sun$,  
with path parameter $\phi \in [0,2\pi]$. 
These paths are closed, starting and ending 
at the identity. As the fundamental group of 
$\sun$ is trivial~\cite{nakahara03:geomet}, 
these paths are homotopic, 
and any homotopy between them, 
with $\theta$ as the homotopy parameter 
(for example with $\pi/2 \le \theta \le 3\pi/4$), 
provides the interpolation we need. 

Let then $n$ be even. 
The formulas \eqref{eq:OmegaNS} for $\Omega_N$ and $\Omega_S$ 
again define two closed paths in $G=\sun/\{\pm\Id\}$, 
starting and ending at the identity, 
with path parameter $\phi \in [0,2\pi]$. 
When these paths are lifted from $G$ to its double cover~$\sun$, formulas 
\eqref{eq:OmegaNS} show that each lift starts at $\Id \in \sun$ 
and ends at $-\Id \in \sun$. As the
fundamental group of $\sun$ is trivial, 
these two lifts are homotopic
to each other in~$\sun$, and this homotopy in $\sun$ projects down into a
homotopy between the original closed paths in $G=\sun/\{\pm\Id\}$. 
Hence the homotopy between the closed paths in $G$ 
provides again the interpolation we need. 

Finally, note that for even $n$ a connection in 
the trivial $\sun/\{\pm\Id\}$ bundle lifts into a 
connection in the trivial $\sun$ bundle. Using the 
gauge group $\sun/\{\pm\Id\}$ instead of $\sun$ 
for even $n$ is hence just a presentational 
convenience. 

\subsection{Triviality of the geon bundle 
for the configurations of Subsection \ref{sec:spher-specsols}} 
\label{sec:disc-non-triv}

In this subsection we show that the geons of Subsection 
\ref{sec:spher-specsols} have a trivial gauge bundle. 

We showed in Subsection 
\ref{sec:remov-dirac-string} that the black hole 
bundle $P$ is trivial and we can realise it as  
$P \mdef \M \times G$. In this realisation, the geon bundle $P'$  
is the quotient of $P$ by the $\Z_2$ group of bundle 
automorphisms whose nontrivial element $K$ takes the form 
\begin{align}
  K: \M \times G &\to \M \times G; 
  \notag\\
     (x,h) &\mapsto \bigl(J(x), h\cdot\Xi(x)^{-1}\bigr) , 
\label{eq:K-action} 
\end{align}
where $\Xi: \M \to G$ is the gauge function that compensates for
$J$ in a globally regular gauge. 
The $G$-multiplication denoted by a dot 
is matrix multiplication for odd 
$n$ and matrix multiplication up to overall sign for even~$n$. 

We shall work in the globally regular gauge
that is obtained from the gauge 
\eqref{eq:nessGcondition}
by the procedure of
Subsection~\ref{sec:remov-dirac-string}. In this gauge 
we have
\begin{equation} 
\label{comptransf-reggauge}
\Xi(x) = H(x) \Omega {\bigl[H\bigl(J(x)\bigr)\bigr]}^{-1} 
\end{equation}
for $0<\theta<\pi$, 
where $\Omega$ is given by \eqref{eq:Omega-alphas-solved}
and $H$ was defined in Subsection~\ref{sec:remov-dirac-string}. 
It follows from 
\eqref{eq:Omega-alphas-solved}
and 
\eqref{eq:OmegaNS} 
that 
$\Xi$ takes a constant value in sufficiently small punctured 
neighbourhoods of $\theta=0$ and $\theta=\pi$. 
$\Xi$ is therefore well defined on $\M$, 
by 
\eqref{comptransf-reggauge} for 
$0<\theta<\pi$ and by 
continuity at 
$\theta=0$ and $\theta=\pi$. 

Recall that a principal bundle is trivial iff it 
admits a global section. 
The geon
bundle $P'$ admits a global section iff $P$ 
admits a global section $\sigma$ that is invariant under~$K$. 
By~\eqref{eq:K-action}, this invariance condition reads 
\begin{equation}
\label{eq:sigmacondition}
\sigma\bigl(J(x)\bigr) = \sigma(x)\cdot\Xi(x)^{-1}, 
\ \ \ \forall x\in\M. 
\end{equation} 

As the gauge potential depends on $U$ and $V$ 
only through the combination~$UV$, it suffices to 
consider 
the condition \eqref{eq:sigmacondition} on the two-sphere at $U=V=0$. 
It further suffices to consider 
\eqref{eq:sigmacondition} 
on the equator $\theta=\pi/2$ of the two-sphere. To see this, let
$\gamma$ and $\Xi_{\mathrm{eq}}$ denote the respective restrictions of 
$\sigma$ and $\Xi$ to the equator. The restriction of 
\eqref{eq:sigmacondition} to the equator then reads 
\begin{equation}
\gamma(\phi+\pi) = \gamma(\phi)\cdot\Xi_{\mathrm{eq}}(\phi)^{-1}. 
\label{eq:gammacurve}
\end{equation}
If $\sigma$ exists, 
it defines a solution to \eqref{eq:gammacurve} by restriction. 
Conversely, suppose that a solution to \eqref{eq:gammacurve} exists. 
We can view $\gamma$ equivalently as 
a $G$-valued function on $S^1$ or as 
a closed path in~$G$, denoted by the same letter and 
given by  
$\gamma: [0,2\pi] \to G$; 
$\phi \mapsto \gamma(\phi)$. 
When viewed as a closed path, $\gamma$ is contractible. 
For odd $n$ this follows because 
$G = \sun$ has a trivial fundamental group. 
For even $n$ the fundamental group of 
$G = \sun/\{\pm\Id\}$ is~$\Z_2$, but $\gamma$ is contractible 
by the observation made in the last paragraph of 
Subsection~\ref{sec:remov-dirac-string}, 
or alternatively by the explicit construction of 
$\gamma$ below. 
Given~$\gamma$, we can define $\sigma$ for $0\le\theta\le\pi/2$ 
by an arbitrary contraction of $\gamma$ into a trivial path at 
$\theta=0$. Defining $\sigma$ for 
$\pi/2 < \theta \le\pi$ by 
\eqref{eq:sigmacondition} then gives the desired~$\sigma$. 

What hence remains is to show that a 
solution to \eqref{eq:gammacurve} exists. 
We now proceed to construct such a solution. 

Let $H_{\mathrm{eq}}$ denote the restriction of $H$ to the equator. 
From \eqref{comptransf-reggauge} we have 
$\Xi_{\mathrm{eq}}(\phi) = H_{\mathrm{eq}}(\phi) 
\Omega {\bigl[H_{\mathrm{eq}}(\phi+\pi)\bigr]}^{-1}$. Defining 
\begin{align}
\label{eq:gammatildedef}
\tilde{\gamma}(\phi)
\mdef \gamma(\phi) \cdot H_{\mathrm{eq}}(\phi+\pi), 
\end{align}
the condition 
\eqref{eq:gammacurve}
can be rearranged into 
\begin{equation}
\tilde{\gamma}(\phi+\pi) = \tilde{\gamma}(\phi)\cdot\Omega^{-1}. 
\label{eq:gammatildecurve}
\end{equation}
Without loss of generality, we may set $\tilde{\gamma}(0) = \Id$; then
$\tilde{\gamma}(\pi) = \Omega^{-1}$. 

Since $\Omega^{-1}$ is special
unitary, it can be diagonalised by 
\begin{equation}
  \Omega^{-1}=UDU^{-1},
\label{eq:Omega-diag}
\end{equation}
where $U$ is unitary and $D$ is a diagonal 
special unitary matrix whose 
diagonal elements are the 
eigenvalues of~$\Omega^{-1}$. 
We need to analyse these eigenvalues. 

Let $n$ be odd. 
A~recursive evaluation of the determinant shows that 
$\left|\Omega^{-1} - \lambda \Id\right| = 
- {(\lambda^2 -1)}^{(n-1)/2} \bigl(\lambda - {(-1)}^{(n-1)/2}\bigr)$. 
The eigenvalues of $\Omega^{-1}$ are hence~$\pm1$, 
and the multiplicity of $-1$ is even. 
We now define the $\phi$-dependent matrix $\Dhat(\phi)$
by replacing an arbitrarily-chosen 
half of the $-1$s in $D$ by $e^{i\phi}$ 
and the other half by 
$e^{-i\phi}$. It is immediate that 
$\Dhat(\phi) \in G$, $\Dhat$ has period $2\pi$, 
$\Dhat(0) = \Id$ and 
$\Dhat(\pi) = D$. 
Given~$\Dhat$, we define 
$\tilde{\gamma}(\phi) \mdef U \Dhat(\phi) U^{-1}$. Then 
$\tilde{\gamma}(\phi) \cdot \Omega^{-1} = U \Dhat (\phi) U^{-1} U D
U^{-1} = U \Dhat (\phi) D U^{-1} = U \Dhat (\phi) \Dhat(\pi) U^{-1} =
U \Dhat (\phi+\pi) U^{-1} = \tilde{\gamma}(\phi+\pi)$, 
so that 
$\tilde{\gamma}$ satisfies \eqref{eq:gammatildecurve}
and $\gamma$ satisfies~\eqref{eq:gammacurve}. 

Let then $n$ be even. 
Proceeding as above, we find 
$\left|\Omega^{-1} - \lambda \Id\right| 
= {(\lambda^2+1)}^{n/2}$. The eigenvalues of $\Omega^{-1}$
are hence~$\pm i$, each with multiplicity $n/2$. 
We now define $\Dhat(\phi)$ by replacing in 
$D$ the eigenvalues $i$ by $e^{i\phi/2}$ and the eigenvalues 
$-i$ by~$e^{-i\phi/2}$. Then $\Dhat(0) = \Id$, 
$\Dhat(\pi) = D$,
and although $\Dhat$ is not $2\pi$-periodic as an $\sun$ matrix,
it is as a $G = \sun/\{\pm\Id\}$ matrix. 
Defining again 
$\tilde{\gamma}(\phi) \mdef U \Dhat(\phi) U^{-1}$, the conditions 
\eqref{eq:gammatildecurve}
and \eqref{eq:gammacurve} can be verified as for odd~$n$. 

Finally, for even~$n$, we verify explicitly 
the claim that 
the path 
$\gamma: [0,2\pi] \to G$; 
$\phi \mapsto \gamma(\phi)$ constructed above 
is contractible in~$G$. Without loss of generality, 
the gauge function $H$ can 
be chosen to equal 
$\Omega_N$
\eqref{eq:OmegaN} on the equator. In this gauge 
it is transparent that the lift of 
$H_{\mathrm{eq}}$ into $\sun$ 
is periodic in $\phi$ with period $4\pi$ 
and changes sign after a translation in $\phi$ by $2\pi$. 
From \eqref{eq:gammatildedef} 
it follows that 
the lift of $\gamma$ to $\sun$ is a 
closed path in~$\sun$, and the contraction 
of this lift in $\sun$ projects down to a 
contraction of $\gamma$ in~$G$. 

This completes the proof of triviality of the geon bundle.

\subsection{Generic gauge field configurations: geon quotient with 
gauged charge conjugation}
\label{sec:spher-gensol-enlarged}

We saw in Subsection \ref{sec:spher-specsols}
that a geon quotient with gauge group $G$ 
does not exist for generic gauge field configurations. 
A~similar obstacle 
for the Maxwell gauge field in the 
Reissner-Nordstr\"om black hole 
\cite{Misner:1957mt} 
can be overcome by promoting $\uone$ charge conjugation 
from a global symmetry into a gauge symmetry~\cite{Louko:2004ej}. 
In this subsection we show that a similar gauging 
of charge conjugation 
works also for the $\sun$ black holes at 
hand. 

In the abelian case, the usual Maxwell gauge group 
$\uone \simeq \sotwo$ is enlarged 
into $\otwo \simeq \Z_2 \ltimes
\sotwo \simeq \Z_2 \ltimes \uone$. 
In the $\Z_2 \ltimes \uone$ 
representation, the 
group multiplication law reads 
\begin{equation}
(a_1,u_1) \cdot (a_2,u_2) = 
\bigl(a_1 a_2, u_1 \rho_{a_1}(u_2)\bigr) , 
\label{eq:enl-multiplication}
\end{equation}
where $a_i \in \Z_2$, $u_i \in \uone$, 
and 
$\rho: \Z_2 \to \Aut\bigl(\uone\bigr), a \mapsto \rho_a$, 
is the group homomorphism 
for which the nontrivial element of $\Z_2$ 
acts on $\uone$ by complex conjugation. Writing 
$\Z_2 \simeq \{0,1\}$, where the identity element is~$0$, 
the explicit formula for $\rho$ is  
\begin{subequations}
\label{eq:rho-def}
\begin{align}
  \rho_0(u) &= u, 
  \\ 
  \rho_1(u) &= \overline{u} . 
\label{eq:rho-def-nontriv}
\end{align}
\end{subequations}

In the nonabelian case at hand, 
the original gauge group $G$ is 
$\sun$ for odd $n$ and $\sun/\{\pm\Id\}$ 
for even~$n$. We enlarge $G$ to 
$\Genl \mdef \Z_2 \ltimes G$ by 
\eqref{eq:enl-multiplication} and~\eqref{eq:rho-def}. 
The group
multiplication table of $\Genl$ reads
\begin{align}
  (0 ,u_1) \cdot (0 , u_2) &= (0 ,u_1 u_2), 
  \notag\\
  (0 ,u_1) \cdot (1 , u_2) &= (1 ,u_1 u_2), 
  \notag\\
 (1 ,u_1) \cdot (0 , u_2) &= (1 ,u_1 \overline{u}_2),  
  \notag\\
  (1 ,u_1) \cdot (1 , u_2) &= (0 ,u_1 \overline{u}_2).  
\end{align}
If $\Omega$ is a gauge function with values in~$G$, 
it follows that the gauge function 
$\widetilde{\Omega}
\mdef (a,\Omega)
: \M \to \Genl$
transforms the gauge potential by 
\begin{equation}
\label{eq:Otildetransf}
  A \mapsto \widetilde{\Omega} A \widetilde{\Omega}^{-1} 
    + \widetilde{\Omega} \d \widetilde{\Omega}^{-1} =
  \begin{cases}
    \Omega A \Omega^{-1} + \Omega \d \Omega^{-1} 
      &\text{if $\widetilde{\Omega}=(0,\Omega)$},\\
    \Omega \bar{A} \Omega^{-1} + \Omega \d \Omega^{-1} 
      &\text{if $\widetilde{\Omega}=(1,\Omega)$}. 
  \end{cases}
\end{equation}

To find a geon, we follow 
Subsection \ref{sec:spher-specsols}
with $\Omega$ 
replaced by~$\widetilde{\Omega}$. 
The conditions \eqref{conditions_genericsols} 
are replaced by 
\begin{subequations}
\label{conditions_genericsols_enl}
\begin{align}
  \widetilde{\Omega} \Lambda_1 \widetilde{\Omega}^{-1} &= -\Lambda_1, 
  \label{eq:17pe}\\
  \widetilde{\Omega} \Lambda_2 \widetilde{\Omega}^{-1} &= \Lambda_2, 
  \label{eq:17e}\\
  \widetilde{\Omega} \Lambda_3 \widetilde{\Omega}^{-1} &= -\Lambda_3. 
  \label{eq:15e}
\end{align}
\end{subequations}
It follows from \eqref{eq:Lambdas} 
and \eqref{eq:Otildetransf} that 
the set \eqref{conditions_genericsols_enl}
is solved by 
$\widetilde{\Omega} = (1, \Omega)$, where 
\begin{align}
\Omega 
&= \diag\left(i^{-n+1}, i^{-n+3}, \ldots, i^{n-3}, i^{n-1}\right)
\notag\\
&= (-i)^{n-1} \diag\left(1,-1,1,-1,\ldots, (-1)^{n-1}\right).
\end{align}
Hence the black hole bundle now admits a geon quotient 
without restrictions on the gauge field configuration. 

If desired, the geon quotient can be 
described as in Subsection~\ref{sec:disc-non-triv}, 
by adopting in the trivial black hole bundle 
$\M \times \Genl$ a globally regular gauge. 
Now, however, the geon bundle is not trivial, 
since the gauge transformation part of the 
bundle map is in the 
disconnected component of~$\Genl$.

\section{Axially symmetric $\sutwo$ black holes}
\label{sec:stat-axially-symm}

In this section we first review the static, axially symmetric 
Einstein-$\sutwo$
black holes discovered by 
Kleihaus and Kunz~\cite{Kleihaus:1997ic,Kleihaus:1997ws}.
For a generalisation to a negative cosmological constant, 
see~\cite{Radu:2004gu}. 
We then give Kruskal-like coordinates that extend the spacetime across the horizon. 

\subsection{The exterior solution of Kleihaus and Kunz}

A static, axially symmetric metric 
can be written in the isotropic coordinates 
$(t,r, \theta, \phi)$ as 
\begin{equation}
\label{metric_axially_symm}
  d s^2 = - f \d t^2 + \frac{m}{f} \d r^2 + \frac{m}{f}r^2 \d\theta^2 
         + \frac{l}{f}r^2 \sin^2\! \theta\, \d\phi^2, 
\end{equation} 
where the positive functions $f$, $m$ and $l$ 
depend only on $r$ and~$\theta$. 
Here $\theta$ and $\phi$ 
are the usual angular coordinates on the 
(topological) $S^2$, with coordinate singularities 
at $\theta=0$ and $\theta=\pi$; for regularity of the spacetime at these coordinate singularities, we need 
$l/m\to1$ as $\theta\to0$ and as $\theta\to\pi$. 
The spacetime is static, with the 
timelike hypersurface-orthogonal Killing vector~$\partial_t$. 
The Killing vector of axial symmetry is~$\partial_\phi$, 
with the symmetry axis at $\theta=0$ and $\theta=\pi$. 

The ansatz for the gauge potential is 
\begin{equation}
  \label{axialYM}
  A = \frac{1}{2er} \Bigl[ \tau^n_\phi \bigl( H_1 \d r +
      \left(1-H_2\right) r \d\theta \bigr) -n \bigl( \tau^n_r H_3 +
      \tau^n_\theta \left(1-H_4\right) \bigr) r \sin \theta \, \d\phi
  \Bigr], 
\end{equation}
where $e$ is the coupling constant, 
the functions $H_i$ depend only on 
$r$ and~$\theta$, and 
\begin{subequations}
\label{eq:sutwofansatz}
\begin{align}
  \tau^n_r &\mdef \sin\theta \cos n\phi \ \tau^x + \sin\theta \sin
    n\phi \ \tau^y + \cos\theta \, \tau^z,\\
  \tau^n_\theta &\mdef \cos\theta \cos n\phi \ \tau^x + \cos\theta
    \sin n\phi \ \tau^y -\sin\theta \, \tau^z,\\
  \tau^n_\phi &\mdef -\sin n\phi \ \tau^x + \cos n\phi \ \tau^y,
\end{align}
\end{subequations}
where $n$ is a positive integer and, 
to conform to the notation of~\cite{Kleihaus:1997ic,Kleihaus:1997ws}, 
$\tau^x$, $\tau^y$ and $\tau^z$ denote respectively the 
Pauli matrices 
$\sigma_1$, $\sigma_2$ and $\sigma_3$~\eqref{eq:paulimatrices}. 
This ansatz is purely magnetic, with no 
term proportional to~$\d t$. 
The ansatz is static, containing no dependence on $t$, and it is axially symmetric, in the sense that the rotation 
$\phi \mapsto \phi +\alpha$ can be undone by 
a gauge transformation with 
$\exp\bigl[-in(\alpha/2)\tau^z\bigr]$. 
With a $2\pi$
rotation in~$\phi$, the ansatz
undergoes a $2\pi n$ rotation in~$\algsutwo$: 
we hence refer to $n$ as the \emph{winding number\/}. 

Finally, we require both the metric and the gauge 
field to be invariant, in an appropriate sense, 
under the north-south reflection $\theta \mapsto \pi-\theta$.  
For the metric the sense is that of isometry, 
implying that 
$f$, $m$ and $l$ are even under $\theta \mapsto \pi-\theta$. 
For the gauge field the sense is 
\cite{Kleihaus:1997ws} that 
$H_1$ and $H_3$ and are odd and 
$H_2$ and $H_4$ are even 
under $\theta \mapsto \pi-\theta$. 

We are interested in solutions to the Einstein-$\sutwo$ 
field equations with a nondegenerate Killing horizon of 
the Killing vector $\partial_t$ 
at $r =r_h>0$. 
The boundary conditions at the horizon and 
at the symmetry axis and 
the integration of the field equations 
into the exterior region $r>r_h$ were 
discussed 
in \cite{Kleihaus:1997ic,Kleihaus:1997ws,Kleihaus:1997mn,%
Kleihaus:1999cc,Kleihaus:1999pi}, 
and numerical evidence was found that solutions exist, including 
solutions that have an asymptotically 
flat infinity at $r\to\infty$. 
The defining properties of the nondegenerate horizon
in the isotropic coordinates of the anzatz 
\eqref{metric_axially_symm}
are $f(r_h,\theta)=0 = f'(r_h,\theta)$ 
and 
$f''(r_h,\theta) > 0$, 
where the prime indicates derivative with respect to~$r$. 
Working in the dimensionless variable $\delta \mdef (r/r_h -1)$, 
it follows that the near-horizon 
Taylor expansions of the 
metric functions and the gauge potential functions begin 
\begin{subequations}
\label{Taylor_first}
\begin{align}
  f(\delta,\theta) &= f_2(\theta) \delta^2 \left[ 1 - \delta +
    \frac{1}{24}\delta^2 F(\theta) + O(\delta^3) \right], 
\\
  m(\delta,\theta) &= m_2(\theta) \delta^2 \left[ 1 - 3\delta +
    \frac{1}{24}\delta^2 M(\theta) + O(\delta^3) \right],\\
  l(\delta,\theta) &= l_2(\theta) \delta^2 \left[ 1 - 3\delta +
    \frac{1}{12}\delta^2 L(\theta) + O(\delta^3) \right] , 
\end{align}
\end{subequations} 
\begin{subequations}
\label{eq:H-taylor}
\begin{align}
  H_1(\delta,\theta) &= \delta \left( 1 - \frac{\delta}{2} \right)
  H_{11}(\theta) + O(\delta^3),\\
  H_2(\delta,\theta) &= H_{20}(\theta) + \frac{1}{4}\delta^2
  H_{21}(\theta) + O(\delta^3),\\
  H_3(\delta,\theta) &= H_{30}(\theta) + \frac{1}{8}\delta^2
  H_{31}(\theta) + O(\delta^3),\\
  H_4(\delta,\theta) &= H_{40}(\theta) + \frac{1}{8}\delta^2
  H_{41}(\theta) + O(\delta^3),
\end{align}
\end{subequations}
where the $O$-terms may depend on $\theta$ 
and the field equations yield various relations among the
coefficient functions~\cite{Kleihaus:1997ws}. 
One of these relations is 
\begin{equation}
  \frac{1}{m_2} \frac{\d m_2}{\d\theta} 
  - \frac{2}{f_2} \frac{\d f_2}{\d\theta} = 0,
\end{equation}
from which it follows that $f_2^2/m_2$ is independent of~$\theta$, 
implying that 
the horizon has constant 
surface gravity~\cite{Kleihaus:1997ws}. 
The gauge potential can further be chosen
regular everywhere, including $\theta=0$ and $\theta=\pi$
\cite{Kleihaus:1999cc,Kleihaus:1999pi}. 
The $\sutwo$ bundle is thus trivial and the 
gauge potential is expressed in a globally regular gauge. 

In the special case $n=1$ the field equations imply that $l=m$, 
$H_1=H_3=0$, $H_2=H_4$ and all the metric and 
gauge potential functions are independent of~$\theta$. 
The metric and the gauge field are then spherically symmetric, 
and the solution reduces to that of 
\cite{Bizon:1990sr,Kuenzle:1990is}.

\subsection{Kruskal-like extension}
\label{sec:extens-two-exter}

A complication with finding Kruskal-like coordinates that 
cover a neighbourhood of the full bifurcate Killing horizon is 
that the null geodesics with constant $\phi$ generically have  
nontrivial evolution in both $r$ and~$\theta$. However, 
because of the discrete isometry $\theta\mapsto \pi-\theta$, 
the submanifold at $\theta=\pi/2$ is totally geodesic, and 
Kruskal-like coordinates that extend this submanifold across the horizon 
can be 
found as in the spherically 
symmetric case of Subsection~\ref{sec:krusk-like-extens}. 
We shall show that the Kruskal-like coordinates adapted to the 
$\theta=\pi/2$ submanifold can be extended to other 
values of $\theta$ to give a 
$C^0$ extension across the horizon. 
This $C^0$ extension will suffice for taking the 
geon quotient in Section~\ref{sec:su2quotient}. 

We start at $r>r_h$ and define the coordinates $(U,V,\theta,\phi)$ by 
\begin{subequations}
\label{eq:55}
\begin{align}
  U &\mdef - \exp \left[ -\alpha \left(t - \int_{r_0}^r \frac{\sqrt{
          m(r,\pi/2)}}{f(r,\pi/2)} \d r \right) \right] , \\
  V &\mdef \exp \left[ \alpha \left(t + \int_{r_0}^r 
      \frac{\sqrt{m(r,\pi/2)}}{f(r,\pi/2)} \d r \right) \right], 
\end{align}
\end{subequations}
where 
\begin{equation}
\label{eq:45}
  \alpha \mdef \frac{f_2(\pi/2)}{r_h \sqrt{m_2(\pi/2)}}
\end{equation}
and $r_0$ is chosen so that 
\begin{align}
  \int_{r_0}^r  \frac{\sqrt{m(r,\pi/2)}}{f(r,\pi/2)} \d r = \frac{1}{\alpha}
  \left[ \ln\delta -\frac{1}{2}\delta +O(\delta^2) \right] 
\end{align}
as $r\to r_h$. 
The region $r>r_h$ is at $U < 0$ and $V>0$, and the Killing horizon is at 
$UV\to0_-$. The metric in the coordinates $(U,V,\theta,\phi)$ reads 
\begin{gather}
\label{eq:40}
\begin{align}
  d s^2 = &f(r,\theta) \frac{1}{2\alpha^2} \frac{1}{UV} \left[
    \frac{m(r,\theta)}{f(r,\theta)^2}\frac{f(r,\pi/2)^2}{m(r,\pi/2)} + 1 \right] \d U\d V 
\notag 
\\[1ex]
     &+f(r,\theta) \frac{1}{4\alpha^2} \frac{1}{(UV)^2}
  \left[\frac{m(r,\theta)}{f(r,\theta)^2}
    \frac{f(r,\pi/2)^2}{m(r,\pi/2)} - 1 \right]
  \left(V^2 \d U^2 + U^2 \d V^2 \right) 
\notag 
\\[1ex]
   &+ \frac{m(r,\theta)}{f(r,\theta)} r^2 \d\theta^2 +
  \frac{l(r,\theta)}{f(r,\theta)} r^2 \sin^2\! \theta\, \d\phi^2,
\end{align}
\end{gather}
where $r$ is a function of $UV$ by 
\begin{subequations}
\begin{align}
  UV &= -\exp \left[ 2 \alpha \int_{r_0}^r 
    \frac{\sqrt{m(r,\pi/2)}}{f(r,\pi/2)} \d r \right]  
    \label{eq:46}
    \\
    &= -\delta^2 \left[ 1-\delta+O(\delta^2) \right] , \ \ \ r\to r_h .  
   \label{eq:54}
\end{align}
\end{subequations}
Inverting the near-horizon expansion \eqref{eq:54}
and substituting in \eqref{eq:40} 
yields
\begin{align}
d s^2 &= 
- \frac{1}{\alpha^2} f_2(\theta) \left[1+O(UV)\right] \d U\d V
+ \frac{1}{96\alpha^2} f_2(\theta)
\times 
\notag
\\
&
\hspace{7ex} 
\times 
\left[ M(\theta)-M(\pi/2)
-2 F(\theta) +2 F(\pi/2)
+ O\bigl(\sqrt{-UV}\,\bigr)
\right] 
\left(V^2 \d U^2 + U^2 \d V^2\right)
\notag
\\
& 
\hspace{3ex} 
+\frac{1}{\alpha^2} f_2(\theta) 
\left[1+O(UV)\right] 
\d\theta^2
+ \frac{l_2(\theta)}{f_2(\theta)} 
\left[1+O(UV)\right] 
\sin^2\! \theta\, \d\phi^2 . 
\label{eq:53}
\end{align}
Similarly, the near-horizon expansion of the 
gauge potential 
\eqref{axialYM} reads 
\begin{align}
  A = &\frac{1}{2e} \biggl\{ \tau^n_\phi
  \biggl[ -\frac{1}{2}\bigl(1+O(UV)\bigr)H_{11}(\theta)(V\d U+U\d V)
  \nonumber
  \\[1ex]
  &\hspace{9ex}
  +\bigl(1-H_{20}(\theta)+O(UV)\bigr)\d\theta\biggr]
  \nonumber
  \\[1ex]
  &\hspace{3ex}
  -n \Bigl[\tau^n_r \bigl(H_{30}(\theta)+O(UV)\bigr) + \tau^n_\theta
  \bigl(1-H_{40}+O(UV)\bigr)\Bigr] \sin \theta \, \d\phi \biggr\} .
\label{eq:Asutwonearhor}
\end{align}

The components of the metric \eqref{eq:53} 
and the gauge potential \eqref{eq:Asutwonearhor}
are well defined at the horizon, $UV \to 0_-$, 
but the components of the metric are not 
guaranteed to be differentiable 
because of the $O\bigl(\sqrt{-UV}\,\bigr)$ error term. 
Our Kruskal coordinates therefore give a $C^0$ 
extension of the spacetime into a neighbourhood of the bifurcate Killing 
horizon, but they are not sufficiently regular 
for discussing the field equations across the horizon. 
Coordinates that allow a smooth extension are discussed in 
\cite{Racz:1992bp,Racz:1995nh}, but at the expense of rendering the 
discrete isometry that we wish to utilise less transparent.  
We shall work with the above $C^0$ extension.

\section{Geon quotient of the axially symmetric $\sutwo$ black hole}
\label{sec:su2quotient}

In this section we show that the Kruskal-like $\sutwo$ black hole of
Section \ref{sec:stat-axially-symm} has a geon quotient. As in
Section~\ref{sec:geon-quotient}, quotienting the spacetime manifold
proceeds as taking the $\RPthree$ geon quotient of Kruskal
\cite{Misner:1957mt,giulini-thesis,giulini-multiRP3s,Friedman:1993ty}
and the issues of interest to us arise with quotienting the principal bundle
in which the gauge field lives.

\subsection{Spacetime quotient}

Let $\M$ denote the spacetime manifold of the Kruskal-like ($C^0$)
extension covered by the coordinates $(U,V,\theta,\phi)$, with the
coordinate singularities at $\theta=0$ and $\theta=\pi$ understood to
be handled as in Section~\ref{sec:geon-quotient}. The map $J$ defined
by \eqref{eq:Jspher-def} is an involutive isometry without fixed
points, preserving both space and time orientation and mapping the two
exterior regions of $\M$ to each other. The quotient spacetime $\M' :=
\M / \{\Id, J\}$ is hence a time and space orientable black and white
hole spacetime, its single exterior region is isometric to one
exterior region of~$\M$, and it is foliated by spacelike hypersurfaces
of topology $\RPthree \setminus \{\text{point}\}$ with the omitted
point being at an asymptotically flat spatial infinity. $\M'$~is hence
a topological geon in the sense of Sorkin
\cite{sorkin86,Sorkin:1979ja,friedmansorkin80,friedmansorkin82} and we
may refer to it as a topological geon black hole. 

The isometries of $\M'$ may be discussed as in
Subsection~\ref{sec:spacetime-quotient}. In particular, the Killing
vector $\partial_\phi$ of $\M$ is invariant under $J$ and there is
hence an induced $\uone$ isometry group on~$\M'$, 
with subtleties at the orbits
coming from the subset of $\M$ where $U=V$ and~$\theta=\pi/2$. We 
shall regard $\M'$ as an axially symmetric spacetime despite these 
exceptional orbits. The
isometry properties associated with the Killing vector 
$V \partial_V - U \partial_U$ of $\M$ are as in 
Subsection~\ref{sec:spacetime-quotient}.

\subsection{Principal bundle quotient}

Let $A\ext$ denote the gauge potential on~$\M$, given in the
right-hand-side exterior by \eqref{axialYM} and having the
near-horizon form~\eqref{eq:Asutwonearhor}.  We need to investigate
whether there exists a bundle map that projects to $J$ on $\M$ and
leaves the gauge potential invariant. As in
Section~\ref{sec:geon-quotient}, this reduces to examining whether
$A\ext$ is invariant under $J$ up to a gauge transformation.

From the evenness of the gauge potential functions 
$H_2$ and $H_4$ and the oddness of the gauge potential functions 
$H_1$ and $H_3$ under $\theta \mapsto \pi-\theta$, 
and from the properties of the matrices \eqref{eq:sutwofansatz}
under~$J$, it follows that the cases of odd and even 
$n$ require separate treatment. 

Let first $n$ be odd. 
$A\ext$ is then clearly invariant under~$J$, 
and the bundle map can be chosen to be 
\begin{align}
K_{\text{odd}}: 
\M \times \sutwo 
&\to 
\M \times \sutwo; 
\nonumber
\\
(U,V,\theta,\phi,h) 
&\mapsto
(V,U,\pi-\theta,\phi+\pi, h) . 
\label{eq:Kodd}
\end{align} 
$K_{\text{odd}}$ is involutive, 
and the quotient bundle is 
the trivial $\sutwo$ bundle over 
$\M_g \mdef 
\M/\{\Id, J\}$. 
As the gauge potential is invariant under a gauge transformation by 
$-\Id\in\sutwo$, 
the geon bundle can be alternatively taken to be the 
trivial $\sothree \simeq \sutwo/\{\pm\Id\}$ 
bundle over~$\M_g$. 

Let then $n$ be even, and let $A\ext^J$
denote the pull-back of $A\ext$ by~$J$. 
In the right-hand-side exterior covered by 
the coordinates $(t,r,\theta,\phi)$, 
$A\ext^J$ takes the form 
\begin{align}
A\ext^J
&= \frac{1}{2er}
  \Bigl\{(\tau^x \sin n\phi - \tau^y\cos n\phi) 
  \bigl[H_1\d r + (1-H_2)r\d\theta\bigr]
  \nonumber
  \\[.5ex]
& \hspace{8ex}
- n \bigl[ (-\tau^x\sin\theta\cos n\phi-\tau^y\sin\theta\sin
  n\phi+\tau^z\cos\theta) H_3
  \nonumber
  \\[1ex]
& \hspace{13ex}
+ (-\tau^x\cos\theta\cos n\phi-\tau^y\cos\theta\sin
  n\phi-\tau^z\sin\theta)(1-H_4) \bigr] r\sin\theta \, \d\phi\Bigr\}.
  \label{eq:JstarAext}
\end{align}
Comparison with \eqref{axialYM} shows that 
$A\ext^J$ and $A\ext$ do not coincide. They are however taken 
to each other by  
$(\tau^x, \tau^y, \tau^z) 
\mapsto (-\tau^x, -\tau^y, \tau^z)$, 
which is a gauge transformation: 
defining 
\begin{equation}
g_0 \mdef \exp \left( i \frac{\pi}{2} \tau^z \right)
= \begin{pmatrix} i&0\\ 0&-i \end{pmatrix} 
\in \sutwo , 
\label{eq:g0def}
\end{equation}
we have 
\begin{equation}
A\ext = g_0 A\ext^J g_0^{-1} , 
\label{eq:homtransf}
\end{equation}
and \eqref{eq:homtransf}
is a gauge transformation because the 
inhomogeneous term involving $\d g_0$ vanishes. 
The bundle map can thus be chosen 
to be 
\begin{align}
K_{\text{ev}}: 
\M \times \sutwo 
&\to 
\M \times \sutwo; 
\nonumber
\\
(U,V,\theta,\phi,h) 
&\mapsto
\left(V,U,\pi-\theta,\phi+\pi, h \cdot g_0^{-1}\right) , 
\label{eq:Keven}
\end{align} 
where the dot denotes matrix multiplication in $\sutwo$. 
$K_{\text{ev}}$ generates the 
cyclic group of order four, 
$\bar{\Gamma} 
\mdef \{\Id, K_{\text{ev}}, K_{\text{ev}}^2, K_{\text{ev}}^3\}$, 
and the geon bundle
is the quotient 
$\bigl(\M \times \sutwo \bigr) / \bar{\Gamma}$. 
As the normal subgroup $\{\Id,K_{\text{ev}}^2\}
\subset \bar{\Gamma}$ identifies points 
in $\M \times \sutwo$ by 
the position-independent gauge transformation by 
$g_0^2 = -\Id\in\sutwo$, 
and as this gauge transformation leaves the gauge potential invariant, 
the geon bundle 
can be equivalently presented as a 
$\BbbZ_2$ quotient of the 
trivial $\sothree \simeq \sutwo/\{\pm\Id\}$
bundle over~$\M$. 
Explicitly, we may realise the projection 
$\sutwo \to \sothree$, 
$g \mapsto \hat g$, in the defining 
matrix representations so that  
$g \tau^i g^{-1} = \sum_j {\hat g}^i{}_j \tau^j$. 
Note that $\hat{g}_0 = \diag(-1,-1,1)$. 
The involutive bundle map then reads  
\begin{align}
{\hat K}_{\text{ev}}: 
\M \times \sothree 
&\to 
\M \times \sothree; 
\nonumber
\\
(U,V,\theta,\phi,\hat h) 
&\mapsto
\left(V,U,\pi-\theta,\phi+\pi, \hat h \cdot \hat g_0^{-1}\right) , 
\label{eq:Kevenhat}
\end{align} 
where the dot denotes matrix multiplication in $\sothree$. 

The geon bundle for even $n$ is not trivial. 
To see this, we view the 
geon bundle as the quotient 
$\bigl(\M \times \sothree \bigr) /
\bigl\{ \Id, {\hat K}_{\text{ev}}\bigr\}$. 
Suppose this bundle is trivial. 
Proceeding as in the discussion of  
Subsection \ref{sec:disc-non-triv} 
leading to~\eqref{eq:gammacurve}, 
we see that there then exist
a continuous $2\pi$-periodic function 
$\gamma: \BbbR \to \sothree$ such that 
\begin{equation}
\gamma(\phi+\pi) = 
\gamma(\phi) \cdot \hat{g}_0^{-1}  
\label{eq:sothreegammacurve}
\end{equation}
and the closed path $\gamma_0: [0,2\pi] \to \sothree; \ 
\phi \mapsto \gamma(\phi)$ is contractible. 
We may assume without loss of generality that 
$\gamma(0) = \Id \in \sothree$. The condition 
\eqref{eq:sothreegammacurve}
then implies that $\gamma_0$ is homotopic to the path 
$\gamma_1: [0,2\pi] \to \sothree; \ 
\phi \mapsto \gamma_1(\phi)$, where 
\begin{equation}
  \gamma_1(\phi) \mdef
  \begin{pmatrix} 
    \cos \phi& -\sin \phi& 0\\ \sin \phi& \cos \phi& 0\\ 0& 0& 1
  \end{pmatrix} . 
\end{equation}
But as the lift of $\gamma_1$ to 
$\sutwo$ is not closed, 
$\gamma_1$ is not contractible, and hence neither is~$\gamma_0$. 
This is a contradiction and 
implies that the
assumed triviality of the geon bundle cannot hold.

\section{Conclusions}
\label{sec:conclusions-outlook}

We have shown that the static, spherically symmetric 
$\sun$ black hole solutions
of K\"unzle~\cite{Kuenzle:1991wa,Kuenzle:1994ru} and the 
static, axially symmetric $\sutwo$ black hole
solutions of Kleihaus and Kunz~\cite{Kleihaus:1997ic,Kleihaus:1997ws} 
admit topological geon quotients.  
These constructions extend the family of known 
non-abelian Einstein-Yang-Mills geon-type black holes from the 
static, spherically symmetric $\sutwo$ geon-type black hole 
\cite{Louko:2004ej} to include geons with
a more general Yang-Mills gauge group 
and to geons with less symmetry. 

For K\"unzle's static, spherically symmetric $\sun$ 
black holes with $n>2$, 
we showed that a geon quotient generically requires an 
extension of the gauge group from $\sun$ to 
$\BbbZ_2 \ltimes \sun$, where the nontrivial element of 
$\BbbZ_2$ acts on $\sun$ by complex conjugation. 
This means that the 
$\sun$ charge conjugation must be treated as a gauge symmetry, 
rather than just as a global symmetry. 
This gauging is very similar to the 
$\text{U}(1)$ charge conjugation gauging that is necessary 
for  taking a geon quotient of the 
Reissner-Nordstr\"om black hole~\cite{Louko:2004ej}. 
By contrast, 
static, spherically symmetric $\sutwo$ black holes 
were known to
admit a geon quotient without the need to gauge the 
$\sutwo$ 
charge conjugation~\cite{Louko:2004ej}, 
and we showed that the same holds for the 
static, axially symmetric $\sutwo$ black holes of 
Kleihaus and Kunz~\cite{Kleihaus:1997ic,Kleihaus:1997ws}. 

In the cases where gauging the charge conjugation is not required, 
we showed that the geons built from K\"unzle's black holes 
have a trivial gauge bundle, 
whereas those built from the black holes of 
Kleihaus and Kunz have a trivial (respectively nontrivial) 
gauge bundle for odd (even) winding number of the 
gauge field configuration. 
We have not investigated whether this phenomenon 
reflects some deeper geometric property. 

Our results on the axially symmetric solutions 
have a technical limitation in that 
the extension across the 
Killing horizon was $C^0$ but was not guaranteed 
to be differentiable. 
We suspect that this limitation 
is an artefact of a non-optimal 
coordinate choice and 
the results continue to hold within extensions 
of higher differentiability. 
It should be possible to examine this 
question with the techniques of 
R\'acz and Wald~\cite{Racz:1992bp,Racz:1995nh}. 

The topological geon black holes that we have found 
should provide an interesting arena for 
investigating the Hawking-Uhruh effect for quantum fields coupled to 
the background Yang-Mills field. How does the geon's charge 
show up in the Hawking-Unruh effect, 
compared with the Hawking-Unruh effect on the conventional Kruskal-like extension? 
In particular, does the 
Hawking-Unruh effect feel the gauging of $\sun$ 
charge conjugation, 
as it does feel the gauging of 
$\uone$ charge conjugation~\cite{bruschi-MGtalk,BruschiLouko}? 
When the charge conjugation is not gauged, 
does the Hawking-Unruh effect feel the 
triviality versus nontriviality of 
the geon's gauge bundle? 
A~technically simple test
field with which to address these questions might be a multiplet of
charged scalars minimally coupled to the Yang-Mills field. A~more 
interesting case might be a neutrino multiplet, for 
which the additional issue of 
inequivalent spin structures arises~\cite{Langlois:2004fv}.

\section*{Acknowledgements}
We thank Martin Edjvet, Yakov Shnir, Elizabeth Winstanley 
and especially John Barrett 
for helpful discussions. We also thank 
Burkhard Kleihaus and two anonymous referees 
for helpful comments on the manuscript. 
GTK was supported in part by the Sunburst Fund of ETH (Switzerland). 
JL was supported in part by STFC (UK) Rolling Grant 
PP/D507358/1.


\end{document}